\documentclass[superscriptaddress,aps,nofootinbib,
preprintnumbers,twocolumn]{revtex4}
\usepackage{amsfonts}
\usepackage{amsmath}
\usepackage{amssymb}
\usepackage{graphicx}

\begin{document}
\preprint{CERN-TH/2003-153}
\title{A Common Origin for all CP Violations}
\author{G. C. Branco \footnotemark
\footnotetext{On leave of absence at Theory
Division, CERN, Geneva, Switzerland}  }
\affiliation{Departamento de F\'{\i}sica and Grupo Te{\'o}rico 
de F{\'\i}sica de Part{\'\i}culas, 
Instituto Superior T\'{e}cnico, Av. Rovisco Pais, 1049-001 Lisboa,
Portugal.}
\author{P. A. Parada}
\affiliation{Departamento de F\'{\i}sica, Universidade da Beira 
Interior, Rua Marqu\^ es d'\' Avila e Bolama, 6200 Covilh\~ a, 
e Grupo Te{\'o}rico de F{\'\i}sica de Part{\'\i}culas, 
Instituto Superior T\'{e}cnico, Lisboa, Portugal}
\author{M. N. Rebelo \addtocounter{footnote}{-1}\footnotemark}
\affiliation{Departamento de F\'{\i}sica and Grupo Te{\'o}rico 
de F{\'\i}sica de Part{\'\i}culas, 
Instituto Superior T\'{e}cnico, Av. Rovisco Pais, 1049-001 Lisboa,
Portugal.}

\begin{abstract}
We put forward the conjecture that all CP violating phenomena
may have a common origin. In order to illustrate our idea, we 
present a minimal model where CP  is spontaneously broken 
at a high energy scale, through the phase in the vacuum expectation
value of a complex scalar singlet. This single phase is the origin
of both low energy CP violation in the quark and leptonic sectors,
as well as leptogenesis. We also show that in this framework
the strong CP problem may be solved in a simple way through the
introduction of a $Z_4$ symmetry which allows for the implementation
of the Nelson-Barr mechanism.

\end{abstract}

\date{\today}
\maketitle

\section{Introduction}

The phenomenon of CP violation plays a central r\^ ole in Particle
Physics and it has profound implications for Cosmology, since it 
is one of the necessary ingredients \cite{Sakharov:dj}
for generating the observed baryon 
asymmetry of the Universe (BAU). From a phenomenological point of view,
one may consider the following four aspects of CP violation: \\
(i)Quark Sector: CP violation was first discovered in the Kaon sector
about four decades ago \cite{Christenson:fg}
and recently was also detected in the B-sector
through CP asymmetries in neutral B-meson decays
\cite{Aubert:2001sp} \cite{Abe:2001xe} . \\
(ii)Lepton Sector: In the Standard Model (SM), neutrinos are strictly
massless and therefore there is neither leptonic mixing nor CP
violation in the leptonic sector, in the context of the SM. 
However, any extension of the SM which accounts for the 
recently observed neutrino oscillations through nonzero
neutrino masses implies, in general, CP violation in the
leptonic sector which might be detected in future
experiments performed at neutrino factories. 
Forthcoming experiments on neutrinoless double beta decay
may also give indirect evidence for the presence of a non-vanishing
phase in the leptonic mixing matrix.\\
(iii) Generation of BAU: One may also interpret the existence
of a matter dominated Universe as another evidence for CP 
violation. It has been established that within the framework
of the SM it is not possible to generate the observed size of BAU,
due in part to the smallness of CP violation in the SM. This
provides motivation for considering new sources of CP violation
beyond the Kobayashi-Maskawa (KM) mechanism. A very interesting 
scenario for generating BAU is that provided by 
leptogenesis \cite{Fukugita:1986hr}, 
where first the out-of-equilibrium decay of righthanded neutrinos
creates a lepton asymmetry which is then converted into
a baryon asymmetry through B-violating but (B-L) conserving
sphaleron mediated processes \cite{Kuzmin:1985mm}. 
An interesting question is
whether, whithin the framework of leptogenesis, it is possible
to relate the CP violation necessary to generate BAU, with leptonic 
CP violation observable at low energies \cite{varios}, 
\cite{Branco:2001pq},  \cite{Rebelo:2002wj}. 
It has been shown 
that this connection exists only in 
specific models \cite{Branco:2001pq}, \cite{Rebelo:2002wj}. \\
(iv) The Strong CP problem: Another aspect of CP violation has 
to do with the fact that in the context of the SM and taking into 
account nonperturbative instanton effects, strong interactions do 
violate CP. This leads to the so-called strong CP 
problem \cite{'tHooft}
for which various solutions have been proposed. \\
\indent
In this paper we address the question of whether it is possible
to find a framework where all these manifestations of CP violation
have a common origin. In particular, we describe a minimal model
with spontaneous CP violation, where CP breaking both in the
quark and leptonic sectors arises solely from a phase $\alpha$
in the vacuum expectation value of a complex scalar singlet
S, with $\langle S \rangle = \frac{V}{\sqrt{2}} \exp (i \alpha )$. 
Since S is an $ SU(2) \times U(1) \times SU(3)_c $ singlet, 
$V$ can be much larger
than the electroweak breaking scale. Therefore, in this framework
CP violation is generated at a high energy scale. In order for
the phase $\alpha$ to generate a non-trivial phase at low
energies in the Cabibbo-Kobayashi-Maskawa (CKM) matrix, one
is led to introduce at least one vector-like quark, whose
lefthanded and righthanded components are singlets under $SU(2)$.
In the leptonic sector, righthanded neutrinos play the r\^ ole
of the vector-like quarks, establishing the connection between
CP breaking at high and low energies. 
With the introduction of 
a $Z_4$ symmetry in the Lagrangean it is possible to find 
a solution to the strong CP problem, of the type proposed
by Nelson \cite{Nelson} and 
Barr \cite{Barr:qx}.
We show that in the 
leptonic sector, one can get CP violation required
to have a viable leptogenesis while also generating CP 
violation at low energies, detectable 
for instance through neutrino
oscillations.

\section{The Model}

We add to the SM the following fields: a singlet charge $- \frac{1}{3}$
vectorial quark $D^0$, three righthanded neutrino fields $\nu_{R}^0$
(one per generation) and a neutral scalar singlet field, $S$. We impose
a $Z_4$ symmetry, under which the fields transform in the following
manner:
\begin{eqnarray}
D^0 \rightarrow -D^0, \ \ \ S \rightarrow -S   \nonumber \\
{\psi _ l^0} \rightarrow i \psi _ l^0, \ \ \  
e_R^0 \rightarrow i e_R^0,\ \ \   
\nu_{R}^0 \rightarrow i \nu_{R}^0
\label{zds}
\end{eqnarray}
where ${\psi _ l}^0$ denotes the lefthanded lepton doublets, while
$e_R^0$, $\nu_{R}^0$ stand for the righthanded charged lepton and neutrino
singlets, respectively. All other fields remain invariant 
under the $Z_4$ symmetry. Furthermore, we impose CP invariance on
the Lagrangean, thus constraining all Yukawa couplings to be real. 
In any weak basis (WB) the Yukawa terms can be written as:
\begin{eqnarray} 
{\cal L}_Y = {\cal L}_q + {\cal L}_l  \label{yql}\\
{\cal L}_q = {\overline {\psi _ q^0}} G_u \phi \ u_R^0 +
{\overline {\psi _ q^0}} G_d  \tilde{\phi } \ d_R^0 + \nonumber  \\
+ ({f_q} S + {f_q}^{\prime} S^\ast ) {\overline {D_L^0}} d_R^0 + 
\tilde{M} {\overline {D_L^0}} D_R^0 + h. c.
\label{lq} \\
{\cal L}_l = {\overline {\psi _ l^0}} G_l \phi \ e_R^0 +
{\overline {\psi _ l^0}} G_{\nu}  \tilde{\phi } \ \nu _R^0 +  \nonumber \\
\frac{1}{2} {\nu} _R^{0T} C ({f_\nu} S + 
+ {f_\nu}^{\prime} S^\ast )\nu _R^0 +  h. c.   
\label{ll}
\end{eqnarray}
Here $\psi_q^0$, $u_R^0$, and $d_R^0$ are the SM quark fields, and
$\phi$ is the SM Higgs doublet.
Notice that an additional bare mass term  of the form
$ \tilde{M} {\overline {D_L^0}} D_R^0$
was included in ${\cal L}_q$. This term is
both gauge and  $Z_4$ invariant and is present in
the Lagrangean together with 
the mass terms arising from the Yukawa interactions
upon $ SU(2) \times U(1) \times Z_4 $ symmetry breaking.
The scalar potential will contain terms in $\phi$
and S with no phase dependence, together with terms of 
the form \mbox{$({\mu }^2 + \lambda_1
 S^\ast S +\lambda_2 {\phi ^ \dagger } \phi )(S^2 + S^{\ast 2}) +
\lambda_3 (S^4 + S^{\ast 4})$} which, in general, lead to 
the spontaneous breaking of T and CP invariance  \cite{Bento:1990wv}
with $\phi$ and $S$ acquiring vacuum expectation values (vevs) 
of the form:
\begin{equation}
\langle {\phi}^0 \rangle = \frac{v}{\sqrt 2}, \ \   \ \ \   
\langle S \rangle = \frac{V \exp (i \alpha )}{\sqrt 2}
\label{vev}
\end{equation}

\section{The Hadronic Sector}
A crucial aspect of this model is the fact that the phase
$\alpha \equiv \arg \langle S \rangle $ arising at a high
energy scale does generate at low energies a CP violating phase
$\delta _{KM}$ in the $3 \times 3$ sector of the mixing matrix
connecting standard quarks. In this respect, the presence of
the vector-like quark $D^0$ plays a crucial r\^ ole, since 
it is through the couplings 
$({f_q} S + {f_q}^{\prime} S^\ast ) {\overline {D_L^0}} d_R^0$
that the phase $\alpha$ appears in the effective mass matrix 
for the down standard-like quarks. Without loss
of generality, one may choose to work in a weak basis where
the up quark mass matrix is diagonal. In this basis, 
it can be readily shown \cite{Bento:ez}  that
the $3 \times 3$ $V_{CKM}$ matrix, mixing the standard quarks 
in the charged weak current 
is obtained through the following relations:
\begin{equation}
{V_{CKM}}^{-1} \ h \ V_{CKM} = d^2
\label{vckm}
\end{equation}
\begin{equation}
h \equiv m_d^0 {m_d^0}^\dagger -
(m_d^0 {M_D}^\dagger M_D \  {m_d^0}^\dagger) / \overline{M} ^2
\label{mem}
\end{equation}
where  $d^2 = \mbox{diag} (m_d^2, m_s^2, m_b^2)$, 
$m_d^0 = \frac{v}{\sqrt 2} \ G_d $, $\overline{M} ^2 = M_D 
{M_D}^\dagger + \tilde{M}^2$ and
$M_D = \frac{V}{\sqrt 2} (f_+^q \cos (\alpha) +
i f_-^q \sin (\alpha))$, with $f_{\pm} \equiv f_q \pm  
{f_q}^{\prime} $.

It is clear from Eqs. (\ref{vckm}), (\ref{mem}) that the phase 
$\delta _{KM}$, generated through spontaneous CP violation
is not suppressed by factors of 
$\frac{v}{V}$. Note that we are assuming that the mass terms
$(M_D)_j$ are of the same order of magnitude as $\tilde{M}$.
This is a reasonable assumption since both terms are
$ SU(2) \times U(1) \times SU(3)_c $ invariant. 
For very large $V$ (e.g. $V$ 
$\sim M_{GUT}\sim 10^{15} $ Gev), $\delta _{KM}$ is the only leftover 
effect at low energies,
from spontaneous CP breaking at high energies. For not so
large a value of $V$ (e. g., $V$ of the order of a few Tev) the
appearance of significant flavour changing neutral currents
(FCNC) in the down quark sector leads to new contributions to
$B_d - \overline{B_d} $ and $B_s - \overline{B_s} $ 
mixing which can alter  \cite{fcnc} some of the predictions of the
SM for CP asymmetries in B meson decays. These FCNC are
closely related to the non-unitarity of the
$3 \times 3$ CKM matrix, with both effects suppressed by powers of 
$\frac{v}{V}$.

As a result of the $Z_4$ symmetry, this model satisfies the 
Nelson-Barr criteria  \cite{Nelson},  \cite{Barr:qx}
and therefore the $\overline \Theta $ parameter is zero in
tree approximation. Recall that 
the parameter $\overline \Theta $ associated
with strong CP violation can be written as
$\overline \Theta = \Theta _{QCD} + \Theta _{QFD}$, where
$\Theta _{QCD} = {g_s}^2 F  \tilde{F}/32 \pi ^2$, and 
$\Theta _{QFD}= arg(det\,m)$, $m$ denoting the quark mass matrix.
In this model CP is a good symmetry of the Lagrangean, only
spontaneously broken by the vacuum, which implies  
$\Theta _{QCD} = 0$. Furthermore,
$\Theta _{QFD}$ vanishes at tree level \cite{Bento:ez} 
in a natural way so that higher order corrections to 
$\overline \Theta $ are finite and calculable. 
The symmetry $Z_4$ plays a crucial r\^ ole in the vanishing of
the argument of the determinant of the down type quark 
mass matrix ${\cal M}_d$. 
One-loop corrections are  suppressed by 
small Yukawa couplings which is a general
property of this class of models, as pointed out by 
Nelson \cite{Nelson}. A nice
feature of this model is that one loop corrections are
further suppressed by the ratio $v^2 / V^2$  \cite{Bento:ez} .

\section{The Leptonic Sector}

In the leptonic sector, after spontaneous symmetry breakdown, one
obtains from Eq. (\ref{ll}) the following mass terms:
\begin{eqnarray}
{\cal L}_m  &=& -\left[ \overline{{\nu}_{L}^0} m \nu_{R}^0 +
\frac{1}{2} \nu_{R}^{0T} C M \nu_{R}^0+
\overline{l_L^0} m_l l_R^0 \right] + 
{\rm h. c.} = \nonumber \\
&=& - \left[ \frac{1}{2}  n_{L}^{T} C {\cal M}^* n_L +
\overline{l_L^0} m_l l_R^0 \right] + {\rm h. c.}
\label{lm}
\end{eqnarray}
where $m$, $M$ and $m_l$ denote the neutrino Dirac mass matrix,
the right-handed neutrino Majorana mass matrix and the charged
lepton mass matrix, respectively, and
$n_L = ({\nu}_{L}^0, {(\nu_R^0)}^c)$.
In this model we have:
\begin{eqnarray}
{\cal M}= \left(\begin{array}{cc}
0 & m \\
m^T & M \end{array}\right), \ \ m_l = \frac{v}{\sqrt 2} G_l, 
\ \  m =  \frac{v}{\sqrt 2} G_{\nu} \nonumber \\
M =  \frac{V}{\sqrt 2}( f_{\nu}^+ \cos (\alpha) +
i f_{\nu}^- \sin (\alpha) ) 
\label{mmm}
\end{eqnarray}
with $f_{\pm}^{\nu} \equiv f_{\nu}  \pm  
{f_{\nu} }^{\prime}$.  
It is clear that $m_l$ and $m$ are real and $M$ is complex 
and symmetric. 
In the leptonic sector the $Z_4$ symmetry prevents
the existence of a mass term of the form 
$\frac{1}{2} {\nu} _R^{0T} C {\overline  M} \nu _R^0$.
Yet, a term of this form will be generated through the couplings
of $\nu_{R}^0$ to the scalar singlet S, after $Z_4$ breaking.

In the weak basis where $m_l$ is chosen to be diagonal
and real
the light neutrino masses and the low energy mixing are obtained 
from the diagonalization of the effective neutrino mass 
matrix $m_{eff} \equiv - m \frac{1}{M} m^T$:
\begin{equation}
-K^\dagger m \frac{1}{M} m^T K^* =d_{\nu}, \label{14}
\end{equation}
In this weak basis, the $V_{MNS}$ matrix is given by $K$
after eliminating three of its 
factorizable phases. Although $m$ is a real matrix, 
since $M^{-1}$ is
a generic complex symmetric matrix,  $m_{eff}$ is
also a generic complex symmetric matrix.
Therefore $K$ has three CP violating phases,
one Dirac-type and two  Majorana-type. On the other hand, the
heavy neutrino masses are, to an excellent approximation,
the eigenvalues of the matrix $M$. In the WB where both
$m_l$ and $M$ are diagonal and real, the lepton-number asymmetry,
resulting from the decay of a heavy Majorana neutrino $N^j$
into charged leptons $l_i^\pm$ ($i$ = e, $\mu$, $\tau$) is given by
\cite{sym}:
\begin{eqnarray}
A^j = \frac{g^2}{{M_W}^2} \sum_{k \ne j} 
{\rm Im} \left((m^\dagger m)_{jk} (m^\dagger m)_{jk} \right)
\times \nonumber \\
\times \frac{1}{16 \pi} 
\left( I(x_k)+ \frac{\sqrt{x_k}}{1-x_k} \right)
\frac{1}{(m^\dagger m)_{jj}}
\label{rmy}
\end{eqnarray}
with the lepton-number asymmetry from the $j$ heavy Majorana
particle, $A^j$, defined in terms of the family number asymmetry 
$\Delta {A^j}_i={N^j}_i-{{\overline{N}}^j}_i$ by :
\begin{equation}
A^j = \frac{\sum_i \Delta {A^j}_i}{\sum_i \left({N^j}_i +
\overline{N^j}_i \right)}
\label{jad}
\end{equation}
the sum in $i$ runs over the three flavours
$i$ = e $\mu$ $\tau$, $M_k$ are the heavy neutrino masses,
the variable $x_k$
is defined as  $x_k=\frac{{M_k}^2}{{M_j}^2}$ and
$ I(x_k)=\sqrt{x_k} \left(1+(1+x_k) \log(\frac{x_k}{1+x_k}) \right)$.
From Eq.~(\ref{rmy}) it can be seen that the lepton-number
asymmetry is only sensitive to the CP-violating phases
appearing in $m^\dagger m$ in this WB.

We show next that in the present model $m^\dagger m$, in the
WB where $m_l$, $M$ are diagonal real and positive, will
contain in general the CP violation required by 
leptogenesis.
CP violation in the general case where $m_l$, $m$ and $M$ 
are complex has been
discussed in a previous work \cite{Branco:2001pq}. 
It was shown that in the special 
WB where $m_l$ and $M$ are diagonal real and positive, all CP
violating phases appear in the matrix $m$, which is a general
complex matrix, and therefore can be written as $m = W^\dagger d V =
UH$, where in the first equality $W$ and $V$ are general unitary
matrices with $d$ diagonal real and positive, and the second 
equality is the polar decomposition into the product of a unitary
and a hermitian matrix. Three
phases in $U$ can be eliminated and $m$ is left with six
independent phases. Whilst low energy CP 
violation is only sensitive
to the three phases appearing in $V_{MNS}$, leptogenesis
only sees the three phases appearing in 
$m^\dagger m = H^\dagger H$.

In our special framework, in the WB where $m_l$ and $M$ are real 
diagonal and positive, the most general matrix $m$ can be written as 
$m= O^T d T$ with $O$ orthogonal real and $T$ unitary. 
Since three of the factorizable phases in $T$ do not commute
with the matrix $O$, $m$ still has six independent
phases.  
The important point is that in this model 
the product $m^\dagger m = T^\dagger d^{2} T $
is entirely general and therefore one may have CP
violation required by leptogenesis.
It can be readily seen from the definition of 
$m_{eff}$ that in this framework the absence
of CP violation at high energies (i.e., a real matrix T) 
immediately implies no CP violation at low energies 
(i.e., no CP violation in K). On the other hand, 
if T is a complex matrix, in general it is not
possible to have $m_{eff}$ real, thus implying 
CP violation also at low energies.
A distinctive feature of this scenario is the fact that
$(mm^\dagger)$ is a now a real matrix. Note that in
supersymmetric seesaw models the predictions for
Br($l_i \rightarrow l_j \gamma$) are directly related
to the size of $(mm^\dagger)_{ij}$ and are potentially 
large \cite{Casas:2001sr}. Furthermore, it has been shown
\cite{Ellis:2001yz} that in the limit of exactly
degenerate heavy neutrino masses, in the general case,
CP violation in charged lepton flavour violating
processes arises only from the phase contained in 
$(mm^\dagger)$. Since in the present model $(mm^\dagger)$
is real, CP violation in those processes vanishes in this limit.
Similar arguments apply to the electric dipole
moments of the charged leptons which are already strongly
suppressed in the above limit \cite{Ellis:2001yz}.
It is important to emphasize that in the limit of degenerate 
heavy neutrinos  non-trivial phases
can still be generated in $V_{MNS}$. 

\section{Conclusions}
We suggest that all physical manifestations of CP
violation may have a common origin. In order to 
illustrate our conjecture, we have presented a
specific minimal model where CP is spontaneously broken at 
a higher energy scale, through the vacuum expectation value of
one complex scalar singlet. A vector-like quark is added to the 
spectrum of the SM and its couplings to standard quarks 
play a crucial r\^ ole for the generation of a CP
violating phase in the CKM matrix. 
Righthanded neutrinos acquire
complex mass terms through their couplings 
to the complex scalar singlet. 
We have shown that the model has the remarkable feature
that although it has only one fundamental CP violating
phase, there is CP violation at low energies both in the
quark and lepton sectors and furthermore
there is sufficient CP violation in order to have
viable leptogenesis. 

\begin{acknowledgments}
GCB and MNR thank the Theory Division of CERN for warm 
hospitality.
This work was partially supported by
{\em Funda\c{c}{\~a}o para a Ci{\^e}ncia e a Tecnologia} (FCT,
Portugal) through the projects CERN/FIS/43793/2001,
POCTI/36288/FIS/2000, CFIF - Plurianual
(2/91) and by CERN. 

\end{acknowledgments}

\end{document}